\documentclass[aps,prm,twocolumn,showpacs,superscriptaddress]{revtex4-1}
\usepackage{epsfig}
\usepackage{dcolumn}
\usepackage{epstopdf}
\usepackage{graphicx}
\usepackage{ulem}
\usepackage{color}
\usepackage{bm}
\begin{document}
\title{Giant thermoelectric figure of merit in multivalley high-complexity-factor 
%lanthanum oxysulphate 
LaSO}
\author{Roberta Farris}
\altaffiliation{Present address: ICN2, Barcelona, Spain}
\affiliation{Dipartimento di Fisica, Universit\`a of Cagliari, Cittadella Universitaria, I-09042 Monserrato (CA), Italy}
\author{Francesco Ricci} \altaffiliation{Present address: Materials Sciences Division, Lawrence Berkeley National Laboratory, Berkeley, CA 94720, United States.}
\affiliation{Institute of Condensed Matter and Nanosciences (IMCN), Universit\'e Catholique de Louvain, Chemin des \'Etoiles 8, B-1348 Louvain-la-Neuve, Belgium}
\author{Giulio Casu}
\affiliation{Dipartimento di Fisica, Universit\`a of Cagliari, Cittadella Universitaria, I-09042 Monserrato (CA), Italy}
\author{Diana Dahliah}\altaffiliation{Permanent address: Physics Department, An-Najah National University, Nablus, Palestine}
\author{Geoffroy Hautier}\altaffiliation{Present address: Dartmouth College, Dartmouth, CN, USA}
\affiliation{Institute of Condensed Matter and Nanosciences (IMCN), Universit\'e Catholique de Louvain, Chemin des \'Etoiles 8, B-1348 Louvain-la-Neuve, Belgium}
\author{Gian-Marco Rignanese}
\affiliation{Institute of Condensed Matter and Nanosciences (IMCN), Universit\'e Catholique de Louvain, Chemin des \'Etoiles 8, B-1348 Louvain-la-Neuve, Belgium}
\author{Vincenzo Fiorentini}
\affiliation{Dipartimento di Fisica, Universit\`a of Cagliari, Cittadella Universitaria, I-09042 Monserrato (CA), Italy}
\date{\today}

\begin{abstract}
We report a giant  thermoelectric figure of merit $ZT$ (up to 6 at 1100 K) in $n$-doped lanthanum oxysulphate LaSO.  
Thermoelectric coefficients are computed from  ab initio bands  within Bloch-Boltzmann theory in an energy-, chemical 
potential- and temperature-dependent relaxation time approximation. The lattice thermal conductivity is estimated 
from a model employing the ab initio phonon and Gr\"uneisen-parameter spectrum. The main source of the large $ZT$ is 
the significant power factor which  correlates with a large band complexity factor. We also suggest a possible $n$-type dopant 
for the material based on  ab initio calculations.
% PRM sec. M11-A
\end{abstract}
\maketitle

\section{Introduction}

Thermoelectric materials, which convert heat into electricity via the Seebeck effect, have attracted
considerable interest as components of power generation devices. In addition to being scalable and reliable, thermoelectric
generators are silent and require no moving parts. Sufficiently cheap and efficient thermoelectric devices 
therefore have massive potential for waste-heat recovery in a variety of industrial and consumer processes, such as automotive exhausts, home heating, and large-scale commercial processes  \cite{genref}. 
%However, when compared with competing technologies, 
Industrial thermoelectric devices are still comparatively expensive and inefficient, and  are often relegated to niche applications despite their potential. Nonetheless, this  might change in a number of ways; for 
example, a breakthrough material might be be found with  a high conversion efficiency as measured by the figure 
of merit
\begin{equation} {ZT}=\frac{\ \sigma S^2}{\kappa_e +\kappa_L} T,
\label{equno}
\end{equation}
where $\sigma$ is the electrical conductivity, $S$ the Seebeck coefficient, $T$ the temperature, and $\kappa_e$, $\kappa_L$ the electronic and lattice   thermal conductivities.  
It is thus only natural that much research is aimed at finding materials with  large  $ZT$. 

In this framework, the band complexity factor $C$ has emerged as a very relevant parameter
% which has emerged
from a large-scale data-mining study  \cite{gibbs}. Indeed, the power factor PF = $\sigma$$S^2$ appearing in the numerator of Eq.~\ref{equno} exhibits a power law increase PF$\sim$$C^{0.6}$ as a function of $C$.
%which is given by
The latter is actually obtained as
$ C = N_v K_v$, where $N_v$ is the multiplicity, i.e., the number of equivalent extrema within the Brillouin zone, and $K_v$ is a measure of the anisotropy of the relevant band extremum
%
%Assuming the 
%the  band energy vs. {\bf k} is  a quadratic form near the relevant extremum,
%$$
%K_v = \left[\frac{m_g}{m_h}\right]^{\frac{3}{2}}= \left[\frac{\sqrt[3]{m_1m_2m_3}}{(\frac{m_1m_2m_3}{m_1m_2+m_2m_3+m_1m_3})}\right]^{\frac{3}{2}},$$
%
(the ratio of geometric and harmonic averages of the diagonal elements of the  mass tensor to the power of 3/2).  
Thus,  multiple band-structure extrema (each potentially involving multiple bands) and anisotropic masses are expected to be conducive to large power factors.  
A quick evaluation readily shows that the most direct power-factor amplifier is the band-valley multiplicity $N_v$, whereas the  $K_v$ factor  deviates significantly from 1 (isotropic case) only  for  energy surfaces with unusually strong anisotropy. 

In this paper, we study a material with a large complexity factor (mostly due to a  large  $N_v$), from which follow very large PF and predicted $ZT$, despite a not especially favorable lattice thermal conductivity. 
Lanthanum sulfoxide, LaSO, was identified as a candidate by screening a large (so far unpublished) database of calculated band extrema. Methodologically, such  database 
searches for  significant complexity factors (and especially valley multiplicity) appear to provide useful guidance in the quest for thermoelectric materials. 

%
% 4 valli: Nv=4, Kv~1.3; 2 valli Kv=1, Nv=2; totale C=7.2
%
%

%[[0,0.5,0],[0,0,0.5],[0,0.5,0.5],[0,0,0],[0.5,0,0]]

\section{Methods and  results}

\subsection{General}

The ingredients of $ZT$ are the electronic transport coefficients (electrical and electronic-thermal conductivity, Seebeck thermopower) obtained from the electronic structure,  and  the lattice thermal conductivity, which we discuss specifically in Sec.~\ref{latthcond}.

The electronic transport coefficients  are computed from  the ab initio density-functional band structure as a function of temperature and doping in the relaxation-time approximation to the linearized  Boltzmann transport equation, an approach known as Bloch-Boltzmann theory  \cite{allen,bt2}. We use a model of temperature- and energy-dependent  relaxation time described at length in previous papers  \cite{noi,sno}, as implemented in a publicly-available custom code  \cite{sno,thermocasu}. We concentrate on $n$-type doping, although $p$-doping also gives interesting values.

\begin{figure}[ht]
\centerline{\includegraphics[width=0.8\linewidth]{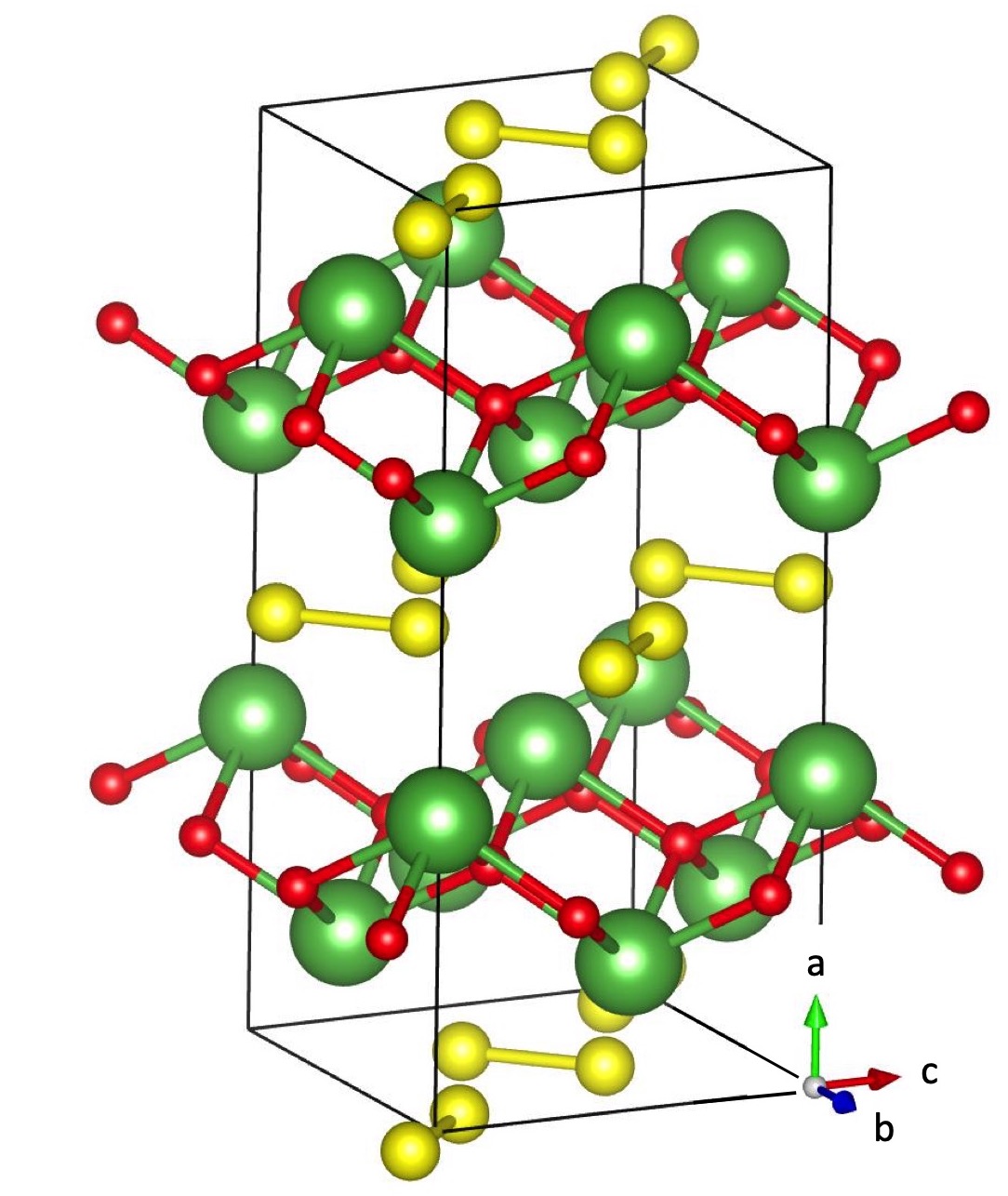}}
\caption{\label{struc} Sketch of the  conventional cell of LaSO (La: large balls; O: small; S: intermediate.}
\end{figure} 

\subsection{Electronic structure calculations}\label{subsec:abInitioCalculations}

Ab initio density-functional structure optimization and band-structure calculations are performed within the generalized-gradient approximation (GGA) using the PBE functional  \cite{pbe} (hereafter simply labelled as GGA) and the projector augmented wave method  \cite{paw} using the VASP code  \cite{vasp}, with the {\tt \verb+La+}, {\tt \verb+S+}, and {\tt \verb+O_s+}  datasets at the maximum suggested cutoff.  LaSO has an orthorhombic structure  (Fig.~\ref{struc}) with space group $Cmca$. It is optimized following quantum forces and stress in a conventional cell containing 8 formula units.  
The material is  made  of strongly buckled  La-O bonded planes intercalated in the $a$ direction by molecular-crystal-like layers of S dimers. The resulting computed lattice constants are $a$=13.323, $b$=5.950, and $c$=5.945 \AA. As expected, transport is found to be less efficient in the $a$ direction than in the $b$$c$ plane. 
The electronic states are calculated   on a (12$\times$24$\times$24) k-points grid.  The minimum gap in GGA is 1.3 eV, so the $n$-type thermoelectric coefficients are essentially unaffected by valence states (we checked this using a scissor operator which adjusts the gap to the value obtained by hybrid functional calculations, see Sec.~\ref{dope}). The  band structure will be discussed further in Sec.~\ref{secbande}.

\subsection{Lattice thermal conductivity}
\label{latthcond}

 We estimate the lattice thermal conductivity $\kappa_L$ using the model of Ref.~\cite{antimonides}. In the present Subsection, the equation numbers refer to that paper. We compute the phonon spectrum and  Gr\"uneisen parameters from first principles using density-functional perturbation theory with the Quantum Espresso  \cite{QE} code, using ultrasoft pseudopotentials and GGA. The phonon-phonon relaxation time is computed from Eqs.3 and 7, where for the parameter $\gamma$ we use the  Gr\"uneisen density of states obtained from the calculated spectrum, instead of the  average Gr\"uneisen value (which we can calculate and find to be 0.78).  Beside phonon-phonon scattering, Casimir and disorder scattering can be easily introduced via Eqs. 21 and 22 (assuming for example that disorder originates from O vacancies).   {The sound velocity is approximated \`a la Debye (constant below the Debye energy $k_B$$\Theta$, zero above); its value is the harmonic average $s$=3134 m/s of the computed small-wavevector acoustic-branch velocities}. The thermal conductivity is finally obtained from Eq.2, with energy dependence only. 

\begin{figure}[ht]
\centerline{\includegraphics[width=1\linewidth]{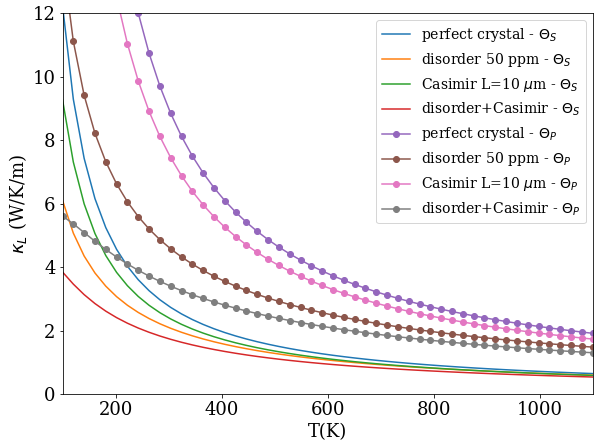}}
\caption{\label{kappaSvsP} Lattice thermal conductivity of LaSO for the perfect crystal, for atomic disorder at 50 ppm (10$^{18}$ cm$^{-3}$), for polycrystalline size 10 $\mu$m, and for disorder plus finite size. The two sets of curves  correspond to  two different definitions  \cite{antimonides,passler}, and hence values, of the Debye temperature: $\Theta_S$=215~K (lines) and $\Theta_P$=335~K (lines with circles).}
\end{figure}

  Fig.~\ref{kappaSvsP} shows the lattice thermal conductivity for two Debye temperatures, $\Theta_S$=215 K  as obtained  {from our data according to} the definition of Eq.~10, and $\Theta_P$=335 K  {with} the definition by P\"assler  \cite{passler}. The choice of $\Theta$ changes considerably the conductivity, because  the phonon-phonon scattering time depends exponentially on $\Theta$. In the following we will compare results obtained with $\kappa_L$ for the perfect crystal  with these two Debye tempeatures (the higher curves for each of the two $\Theta$'s).   Fig.~\ref{kappaSvsP} also shows $\kappa_L$ for a  few  combinations of size and disorder parameters; weak disorder or micron-sized-crystalline Casimir scattering  clearly have comparatively minor effects on the scale set by  the choice of $\Theta$, so we will not expound upon them further.

 \subsubsection{ {Discussion}}
 {The accuracy of the estimate of $ZT$ to be discussed below clearly depends on the lattice conductivity, as a hypothetically much larger $\kappa_L$ would directly reduce $ZT$. A full ab initio computation \cite{noi} is costly and impractical for this material, hence out of our present scope. Nevertheless, we reckon that the model calculation of $\kappa_L$ should be sufficient for the present purposes: it has the correct asymptotic behavior; it includes the density of states and specific heat obtained ab initio, as well as  the ab initio Gr\"uneisen parameters  in the relaxation time;  the only model part is the relaxation time, whereby in particular the key parameter is, as mentioned, $\Theta_D$ for which we use two different possible definitions; finally, the model appears to err on the side of giving too large a $\kappa_L$ \cite{antimonides} compared to experiments and other techniques.}

 {Adding to this the fact that we use the worst possible $\kappa_L$ (that for the crystal) to calculate $ZT$ below, whereas in a real material $\kappa_L$ will be smaller (e.g., defected, doped, polycrystalline, etc.), we ultimately expect no significant $ZT$ reductions compared to our estimates (rather, even an increase). This is because the defining feature of this material is in fact the large electronic power factor.}

\subsection{Transport coefficients and  electronic relaxation time}
\label{trcoe}

   We compute the transport coefficients with a code~\cite{thermocasu,sno} employing parts of the  BoltzTrap2  \cite{bt2} transport code as libraries, and including energy- and temperature-dependent relaxation time. The ab initio bands (assumed rigid, i.e., not changing with doping or temperature) are interpolated by a Fourier-Wannier technique  \cite{bt2} over  a {\it k}-point grid with 64 times more points than the ab initio one, i.e. approximately equivalent to a (48$\times$96$\times$96) grid. 
   
   As discussed for instance in Ref.~\cite{noi}, a constant relaxation time $\tau$=$\tau_0$  neglects relevant physics  {in the electron scattering} (thermal distributions, local masses, energies of phonons, and more). So we adopt a temperature- and energy-dependent relaxation time 
$\tau(T,E,\mu)=1/P_{\rm imp}+1/P_{\rm ac}+1/P_{\rm polar}$ which enters the kernel $\sigma$ of the integral in Eq.~9 of Ref.~\cite{bt2} (additional details can be found in Ref.~\cite{sno}). The $P$'s are   the scattering rates of charged impurities, acoustic phonons, and polar optical phonons given in Refs. \cite{noi,ridley1,ridley2}. Piezoelectric  scattering  is zero by symmetry  \cite{ridley2,bilbao}.  {We point out that there is no effective mass approximation in the transport coefficient calculations, except those implicit in the relaxation time.}
    
The parameters entering the model for $\tau$  can be computed directly.  The conduction-band deformation potential is 
$D$=2.1 eV, the density is 5270 kg$\cdot$m$^{-3}$ and the effective conduction mass is set to a  {conservatively large} $m_c^*$=0.5 $m_e$. The   
phonon energies, dielectric tensor, and sound velocity are obtained from linear-response  calculations  \cite{QE} 
as in Sec.~\ref{latthcond}.   {The sound velocity is 3134 m/s as discussed in the previous Section}. For the dielectric 
constants we use the harmonic averages of the (very similar) $b$ and $c$ components, $\varepsilon_{\infty}$=6.23 
and $\varepsilon_{\rm static}
$=18.73 (as it turns out, the $a$ direction is quite irrelevant for thermoelectricity).

\begin{figure}[ht]
\centerline{\includegraphics[width=1\linewidth]{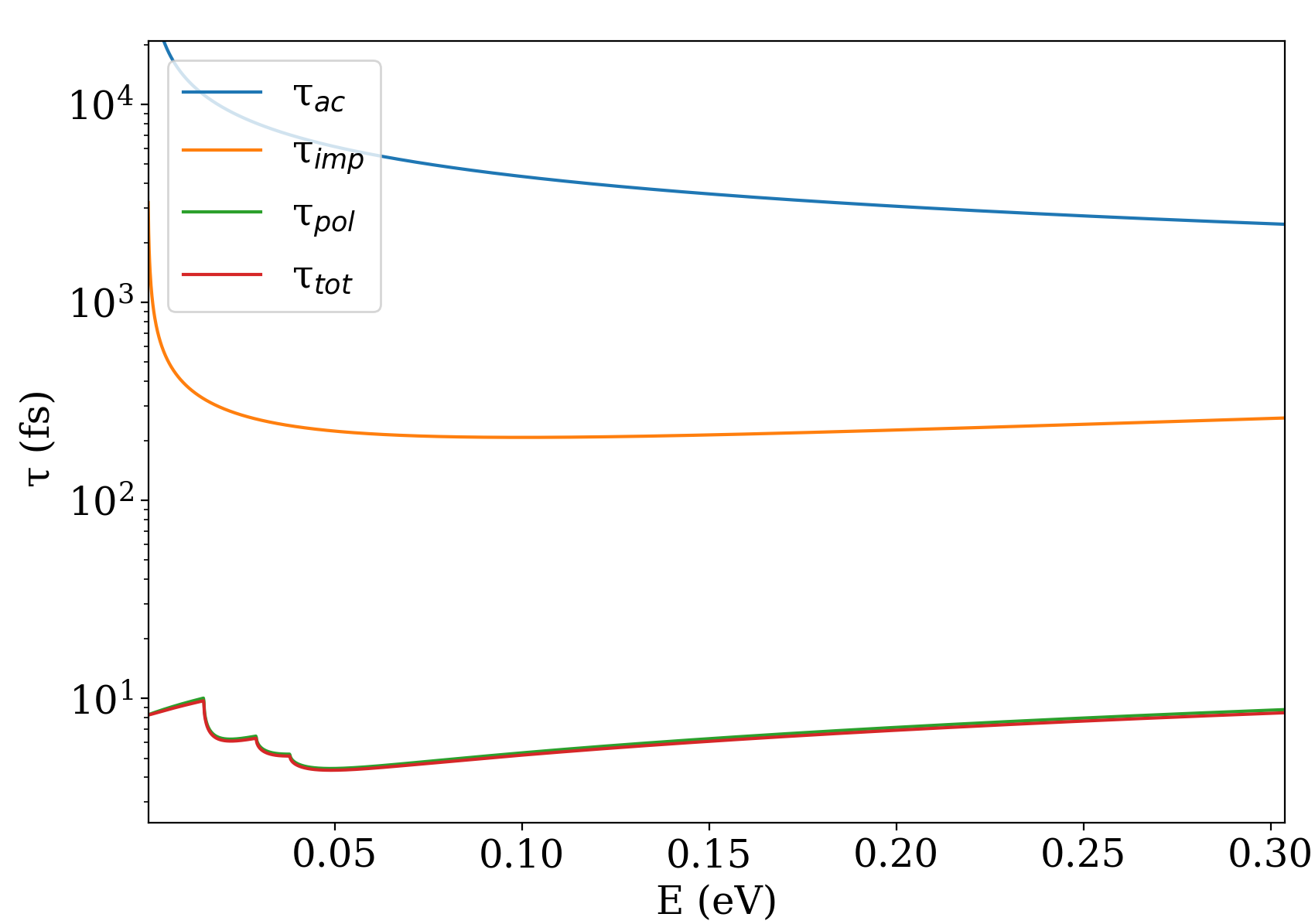}}
\caption{\label{tauE} Relaxation time vs. $E$ for $T$=600 K and $\mu$=0 for the $b$ axis. From top: acoustic-phonon, impurity, polar-phonon, total.}
\end{figure}

A key point is the scattering by polar (i.e. optical) phonons, which dominates the total scattering rate.  The LO phonon energies are calculated directly using the linear-response approach  {(Sec.~\ref{latthcond}) with non-analiticity along the crystallographic directions. Since LO phonons scatter electrons with  momentum parallel to the LO polarization vector,} a direction-dependent polar-phonon $\tau$ would be needed.   {To simplify the procedure, we use} a set of effective LO energies (20, 27, and 30 meV) obtained as  {averages of the three LO groups of frequencies over the $a$, $b$, and $c$ directions}.  {We deem this simplification to be quite acceptable} in view of the other significant uncertainties in the calculations, such as the choice of $\Theta_D$ in the lattice thermal conductivity, or our choice of using a single phonon replica in the polar-phonon scattering rate.

%TO cm-1   55,82,114
%TO meV   6.8, 10.2, 14.1
%LO meV   11.7, 17.6, 24.3, pol=z,y,z

The relaxation time  $\tau$($E$,$T$) is sketched vs. $E$  in Fig.~\ref{tauE}. As typical of polar insulators, the 
polar-phonon scattering dominates, and its downward jump across the LO phonon energies is in the low 
energy  region  relevant  for transport.  Here, we use this relaxation time accounting self-consistently for 
chemical potential changes, and do not employ any  average-time or constant-time approximations.

\subsection{Band structure and complexity factors}
\label{secbande}

\begin{figure}[htbp]
\centerline{\includegraphics[width=1\linewidth]{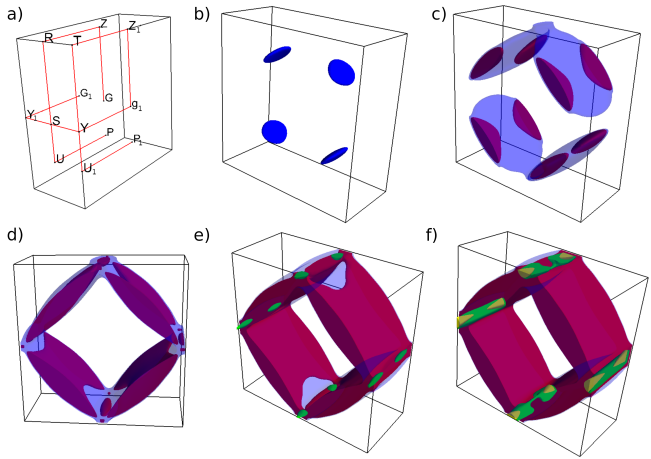}}
\caption{\label{fig:org4228e2d}
a) Brillouin zone with the high symmetry points path used in Fig.~\ref{bandfermidos}. Other panels: Fermi surface at b) 0.01 eV, c) 0.07 ev, d) 0.1 eV, e) 0.11 eV, f) 0.13 eV above the conduction edge. The  conduction bands involved are colored in blue (1st band), red (2nd), green (3rd), yellow (4th).} 
\end{figure}

\begin{figure*}[ht]
%\begin{figure}[ht]
\includegraphics[width=1\linewidth]{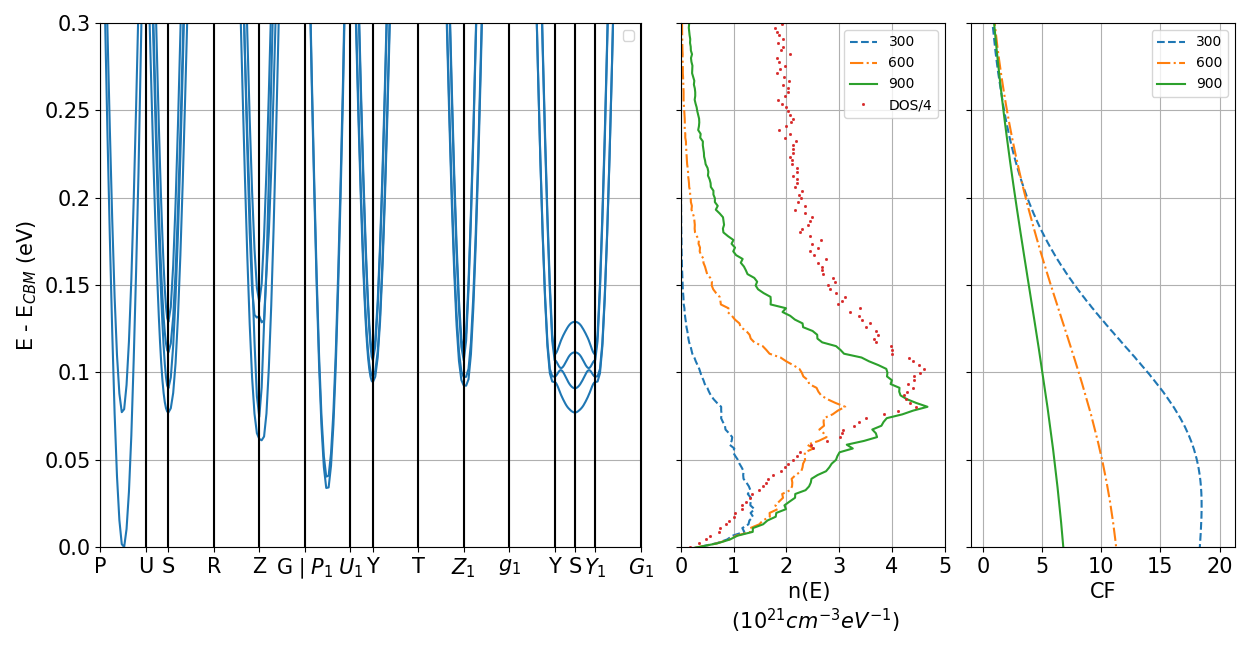}
\caption{\label{bandfermidos} Left: conduction bands of LaSO along the paths marked in Fig.~\ref{fig:org4228e2d}a. Center: the density of states and carrier density (i.e. the conduction density of states multiplied by the Fermi distribution), both in units of 10$^{21}$ cm$^{-3}$ eV$^{-1}$, vs. energy. Right: complexity factor for the $b$ direction ($c$ is similar, $a$ is about zero) vs. energy. The zero of energy is the conduction minimum.
In the center and right panel, $\mu$=0 and $T$=300, 600,  900 K.}
%\end{figure}
\end{figure*}

We now discuss the band structure of LaSO, and two versions of the complexity factor. The first, $C_b$, is a  geometrical value derived from the number of valleys and the masses from the band structure. The second, $C_t$, is a transport value obtained a posteriori from the calculated transport properties, with the procedure of Ref.~\cite{gibbs}. The data are summarized in
Fig.~\ref{fig:org4228e2d}, reporting the conduction-band Fermi surface of $n$-doped LaSO at several chemical potentials, and Fig.~\ref{bandfermidos}, 
displaying the bands, the carrier density (the conduction density of states multiplied by the Fermi distribution), and the complexity factor $C_t$.

%\onecolumngrid
%\twocolumngrid

The low-energy valleys of the LaSO conduction band occur at four internal points of the Brillouin zone (BZ) and at four zone-border points (which count as two internal points). 
By inspection, the first four conduction states (marked in different colors in Fig.~\ref{fig:org4228e2d}) 
provide 16 available valleys 
within about 150 meV of the band edge. These  are visible in the Fermi surface in Fig.~\ref{fig:org4228e2d}.  Four valleys provided by the first band are internal to the BZ. Eight valleys, from the second band, are located on the square faces and therefore shared with adjacent BZs (hence their wight is 1/2). Eight valleys come from the third band (specifically on the segments Z-Z$_1$ and S-Y/S-Y$_1$, Fig.~\ref{bandfermidos}), and eight more from the fourth band on the same segment. 

The same result is found by counting the relevant bands in Fig.~\ref{bandfermidos}. The stationary point on the P--U segment and  its three symmetry partners contribute a total of 8 bands; points S and Z with their two symmetry partners contribute 2 bands each (accounting for their being at zone border);  that amounts again to 16 available band minima.

The carrier density vs. energy (in the central panel of Fig.~\ref{bandfermidos}) shows that all these bands are occupied at the chemical potentials and temperatures of  interest (i.e., $\mu$ near the lowest conduction edge, and temperatures of the order of 1 to 3 times room temperature). Accordingly, all the bands just mentioned  may be considered active (i.e., contributing to the trasport), so  that the effective total  multiplicity of the occupied valleys is $N_v$=16.  From Fig.~\ref{bandfermidos}, we can 
also infer that the optimal doping level, namely the doping at which $ZT$ is maximal for a given $T$, will 
probably fall in the  low 10$^{20}$ cm$^{-3}$, and that the Seebeck coefficient may be interesting due to the fast rise 
of the density of states near the band edge. 

As it turns out, the relevant directions for conduction are in the basal $b$-$c$ plane of the material. The masses in the in-plane directions are quite isotropic, so $K_v$$\simeq$1, and the geometrical complexity factor is $C_b$$\simeq$16. If the  $a$ component of the mass tensor were to be considered,  a larger anisotropy would arise, resulting in $K$$\simeq$1.2-1.3, i.e., a  $C_b$ of about 20. (The anisotropy can be appreciated, for example, from  the different curvature of the  bands  at point S, respectively along the Y-S-Y$_1$ segment and along the U-S-R segment.)

\begin{figure}[ht]
\includegraphics[width=1\linewidth]{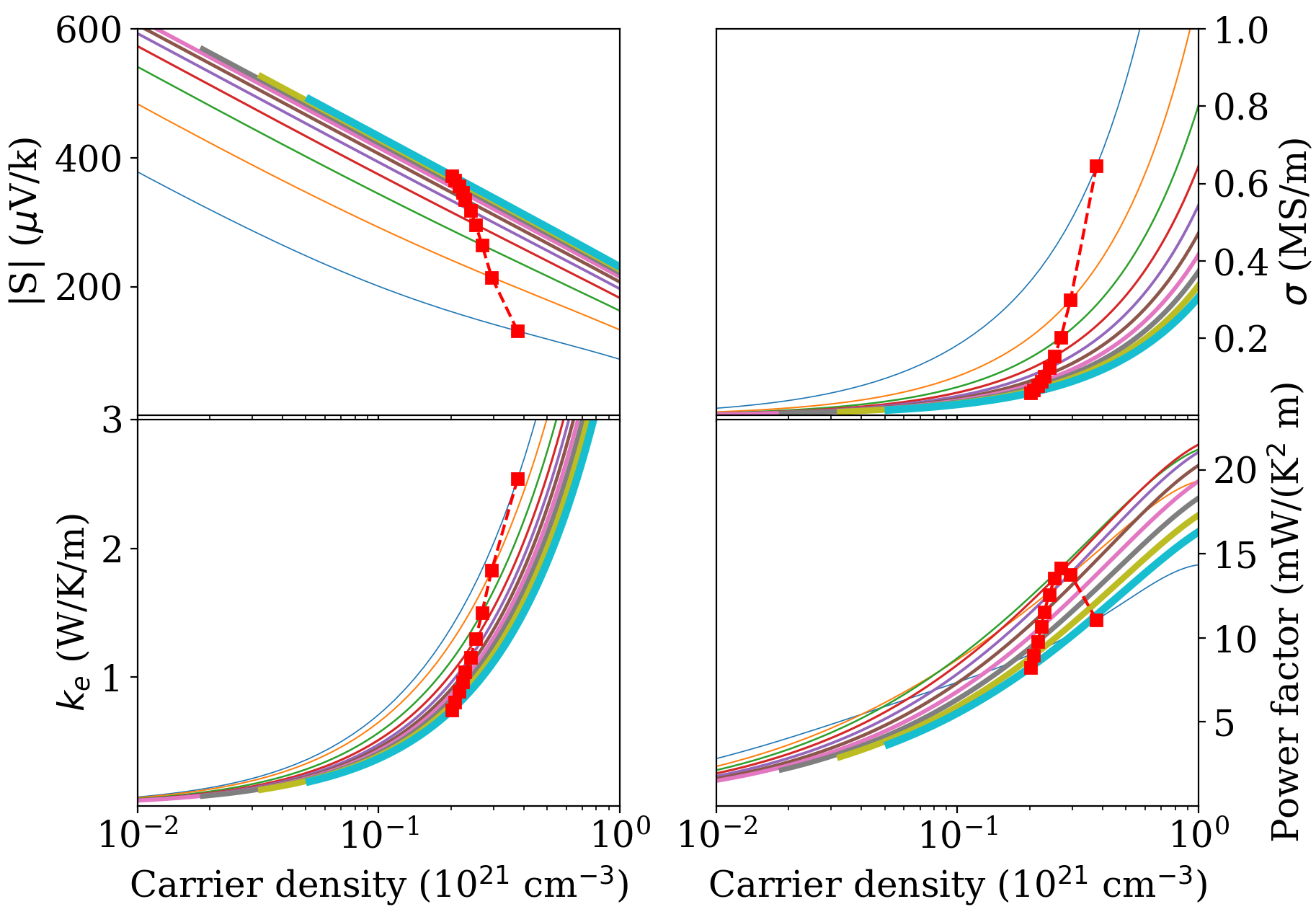}
\caption{Seebeck coefficient, conductivity, electronic thermal conductivity, and power factor  vs. doping  at   {$T$ increasing from 200 to 1100 K, denoted by increasing line thickness}. The red squares are values at optimal doping.}
\label{fig7}
\end{figure}

The  transport complexity factor $C_t$  is in the rightmost panel of Fig.~\ref{bandfermidos}. It  is computed   from the calculated transport coefficients in the basal plane as outlined in Ref.~\cite{gibbs} using a constant relaxation time $\tau_0$=10 fs, purely for consistency with Ref.~\cite{gibbs}. We set the scattering parameter $\lambda$=1 in the Seebeck coefficient model of Ref.~\cite{gibbs}, to take into account the dominant polar scattering we have in the present case. Similarly to most quantities in thermoelectricity, $C_t$ is a temperature- and chemical potential-dependent  tensor. To compare it with the geometric $C_b$ (a scalar) we pick a low T=300 K and $\mu$=0, and average over in-plane directions. $C_t$ at 300 K is relatively flat at low energy, and its energy average is roughly 15, in quite decent agreement with $C_b$=16 obtained above. This is in line with our having considered just the in-plane, largely isotropic transport. We discuss in Sec.~\ref{TE_coeff_sec} (especially with reference to Figs.~\ref{fig7} and \ref{fig8}) the connection of our $C$ values with the rule of thumb of Ref.~\cite{gibbs}.

\subsection{Thermoelectric coefficients and figure of merit}
\label{TE_coeff_sec} 
 
\begin{figure}[ht]
\includegraphics[width=1\linewidth]{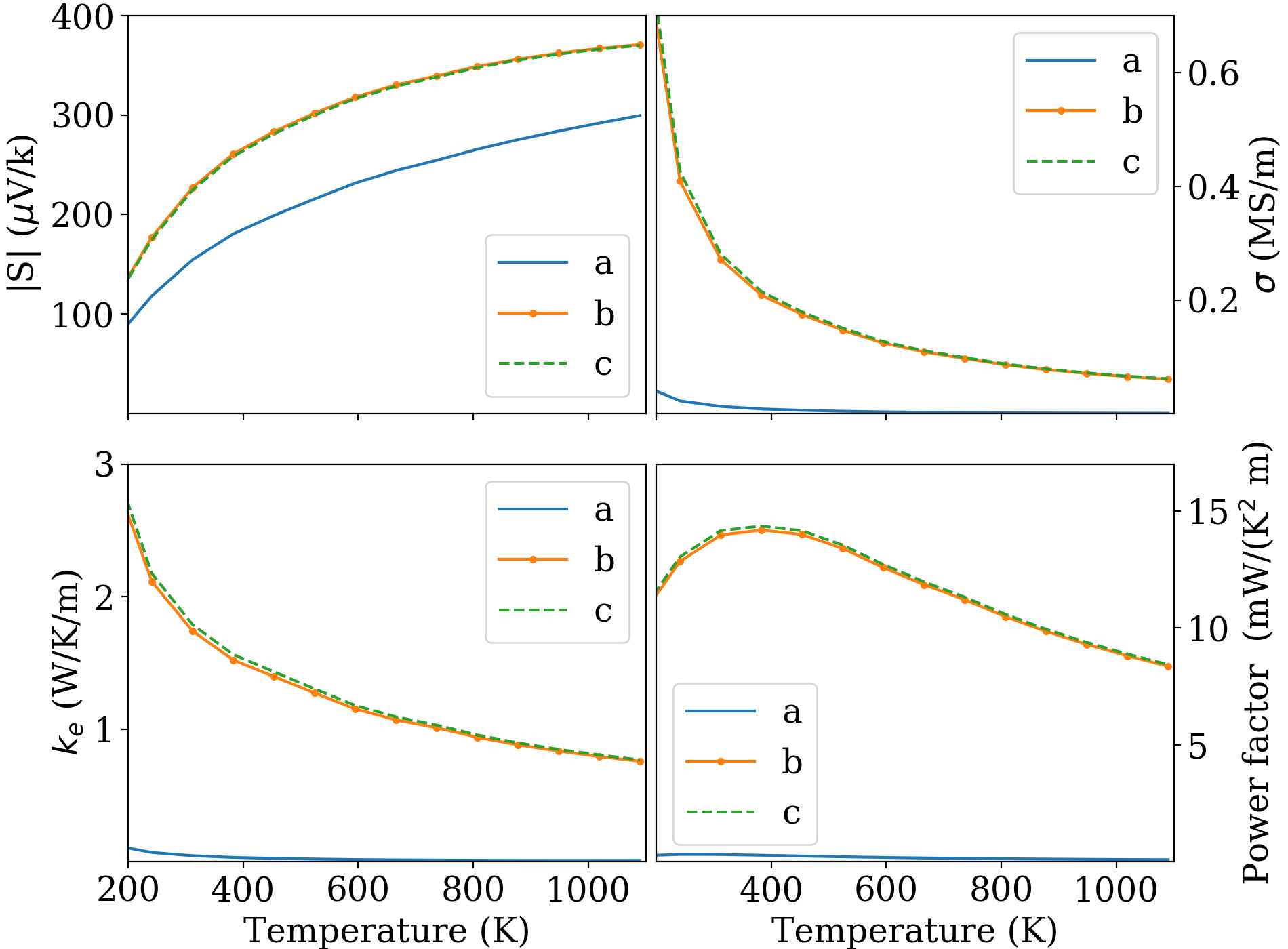}
\caption{Seebeck coefficient, conductivity, electronic thermal conductivity, and power factor  vs. $T$    at optimal doping.}
\label{fig8}
\end{figure}

Fig.~\ref{fig7}  reports the Seebeck coefficient, the electrical conductivity, the electronic thermal conductivity, and the power factor vs. doping and parametrized by $T$. Fig.~\ref{fig8}  displays the same quantities  vs. $T$ at optimal doping (i.e. the doping at which $ZT$ is a maximum at the given temperature). For simplicity, only the $b$ component is plotted in Fig.~\ref{fig7}. As seen in Fig.~\ref{fig8}, the $c$ and $b$ components are very close, and the $a$ component ends up producing small power factor and $ZT$, so it can be ignored.

The large conductivity and Seebeck coefficient result in a large power factor for the in-plane transport, with a maximum of 15 mW/(K$^2$m) at 400 K. We can now make contact  
 between the large  complexity factors $C_t$$\sim$$C_b$$\simeq$16 discussed in the previous Section and the rule of thumb of Gibbs {\it et al.}  \cite{gibbs}. For this complexity factor,
 Fig.~3 of Ref.~\cite{gibbs}  suggests that the expected  maximum power factor should be  between 6 and 22 mW/(K$^2$m). Using the  setting of Ref.~\cite{gibbs} ($\tau$=$\tau_0$=10 fs and $T$=600 K), we indeed 
 obtain a power factor of 21 mW/(K$^2$ m); with the full relaxation time treatment, the power factor  at 600 K (Fig.~\ref{fig8}) is about 12 mW/(K$^2$m). In both cases, our results are consistent with the general prediction   \cite{gibbs}.

\begin{figure}[ht]
\includegraphics[width=1\linewidth]{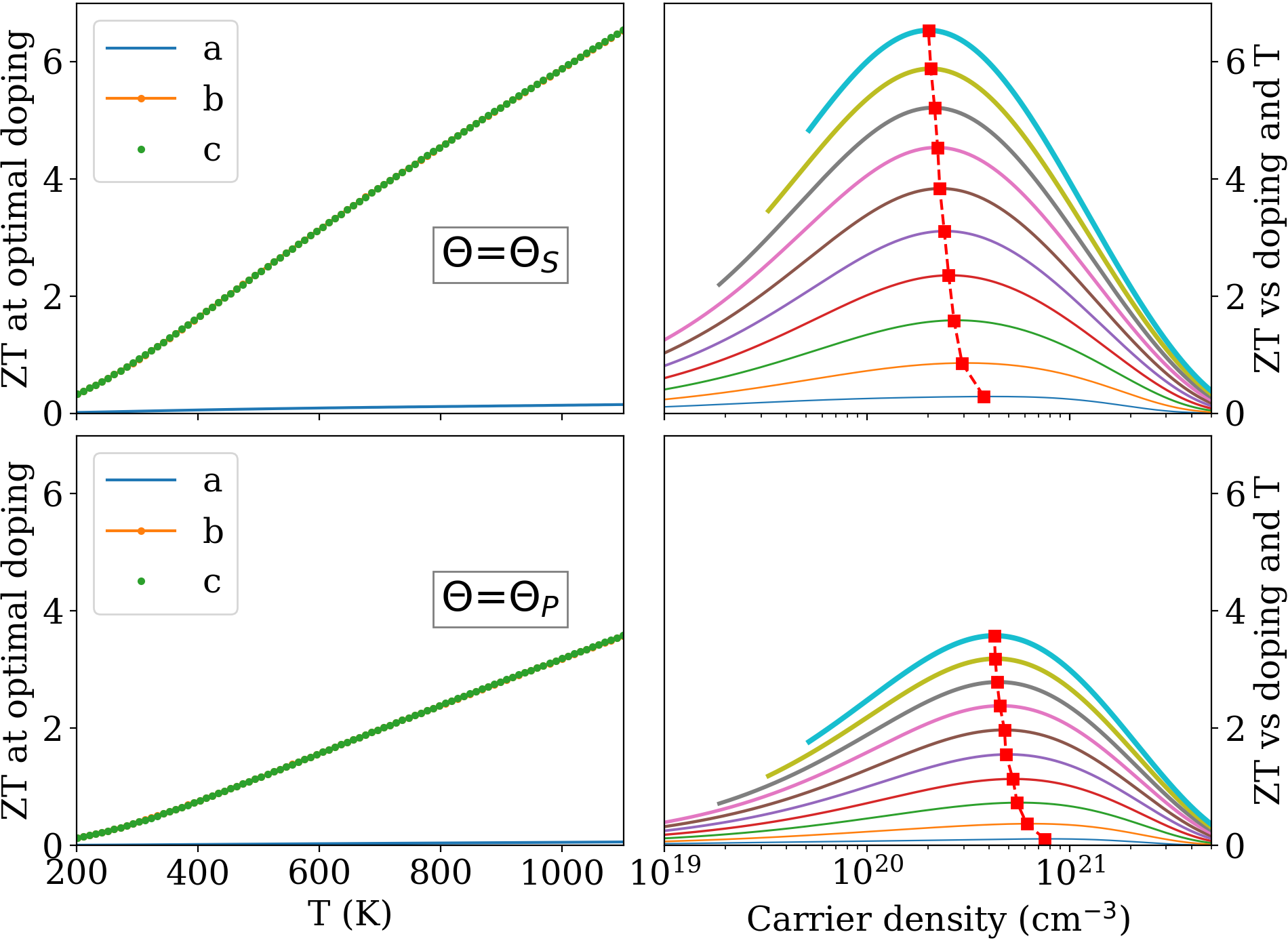}
\caption{\label{fig5}  $ZT$  of LaSO.  Left panels: $ZT$ vs. $T$ at optimal doping;  right panels: $ZT$ vs. doping ($b$ component only) and various $T$. Top panels: lattice thermal conductivity calculated with $\Theta$=$\Theta_S$; bottom panels:  same for $\Theta$=$\Theta_P$. Temperatures 200 to 1100 K, line thickness proportional to $T$. All quantities are drawn on identical scales for easier comparison.  {Note that on this scale the $a$ component is very small, and the $b$ and $c$ component are practically indistinguishable.}}
\end{figure}

Finally, in Fig.~\ref{fig5}, we show the $ZT$ tensor for the two instances of lattice thermal conductivity discussed in Sec.~\ref{latthcond}. The left panels report  the diagonal components of $ZT$  as a function of $T$ at optimal doping, 
%(i.e. the doping at which $ZT$ is a maximum at a given $T$), 
and  the right panels  the $b$ component as a function of doping for  different temperatures.
The $b$ and $c$ components differ by only a fraction of a percent. The main results in this Figure are {\it i)} $ZT$ is very large, reaching values well over 3 and, respectively, 6 at high $T$ for the two versions of the lattice thermal conductivity; {\it ii)}  
the optimal doping is in the low to mid $10^{20}$ cm$^{-3}$, depending on $\kappa_L$ and $T$.  $ZT$ may still be rather interesting even at lower doping: for example, it is already above 2 at 800 K and 2$\times$10$^{19}$ cm$^{-3}$ (Fig.~\ref{fig5}, upper panel).
We recall again that all coefficients refer to electrons in perfect-crystal bands, subject to scattering from phonons and
charged impurities, so that no scattering is accounted for from disorder, dislocations, neutral impurities,  {extended defects}, etc. which could affect transport in ways we cannot quantify.

\subsection{Doping}
\label{dope}

Given the relatively high carrier density required to obtain interesting $ZT$s, we now look into the possibility of $n$-type doping of LaSO.  {Screening a number of  options, we find that Hf seems to be a reasonable candidate donor. }

 {Dopability is difficult to assess in any  material, and LaSO is no exception. In particular, here we do not address possible compensation by native defects or other contaminants, but only the solubility and ionization of donors. Also,} our discussion is based on equilibrium thermodynamics, so the possibility remains that epitaxial growth (which often occurs out of equilibrium), ion implantation, and certain kinds of diffusion doping processes may do better than we predict here. 

Solubility and carrier concentrations are estimated from the formation energies and thermal ionization levels obtained via ab initio calculation. We use  VASP, on 144-atom supercells with a 4$\times$4$\times$4 k-point grid and the GGA, the details being the same as in the calculations in the previous Sections. 
%As discussed further below, we checked our results with the hybrid HSE  functional. 
For charged states we use the simple monopole correction by Leslie and Gillan  \cite{LG}, with the dielectric constant calculated in Sec.~\ref{trcoe}. 

We concentrate on potential donors substituting on the La site (namely Zr, Hf, Ce, Sb, Bi, Sn, and Si) as we find that 
potential  substitutions for O or S, such as F or Cl, tend to go interstitial or have large formation energy.  Si can be 
discarded offhand because of its huge formation energy. Sb, Bi, and Sn can be neglected as well since they have no 
state (in particular no donor state) in the vicinity of the gap. This effective trivalent behavior is presumably due to the
hybridization cost of the $s$ orbital, and the pyramidal bonding geometry of the La site which does not favor $sp$ 
hybrids.

\begin{table}[ht]
\caption{Donor thermal levels (eV from conduction edge).}
\label{tab}
 \begin{tabular}{@{} lcccc @{}}
 \hline   
 Dopant& Hf & Zr & Zr HSE & Ce \\
 $\varepsilon$ & +0.05 & +0.05 & +0.07  & --0.56 \\
\hline
\end{tabular}
\end{table} 
For the remaining dopants Zr, Hf, and Ce, we find the thermal levels reported in Table \ref{tab}. Neutral Ce has an electron in an $f$ orbital when substituting for trivalent La, and its donor level  is deep in GGA. This will not improve when  using 
non-local-density methods (such as hybrid functionals) which remove self-interaction and tend to lower the energy of very localized occupied states. Luckily, instead, Zr and Hf have shallow thermal levels, in fact lying just above the 
conduction edge in GGA (although there as usual are large uncertainties in these estimates, at the very least 
$\pm$0.1-0.2 eV).   {A python notebook with the formation energies and their post processing is at {\tt https://gitlab.com/vfiore/laso-eform}.}

To check for the effects of more advanced exchange-correlation functionals, we calculated the 
thermal level for Zr using the  hybrid HSE functional  \cite{hse}, and found a value similar to GGA. This  is further 
confirmed by the electronic band structure, which exhibits impurity-related resonant states within the lower portion 
of the conduction band, at the same position both in GGA and HSE. In passing, the gaps are 1.3 eV in GGA and 2.9 eV
and HSE, the difference being within 15\% of that predicted by the dielectric correction rule  \cite{FB}. 

Another 
interesting result of these calculations is 
that the upper valence band and the bottom conduction band  are  predominantly sulphur-like, so that 
conduction effectively occurs in-plane in the S layers. Carriers living in the S layer may thus be partially decoupled from charged impurities in the La-O layers, leading to a reduced impurity scattering.

\begin{figure}[ht]
\centerline{\includegraphics[width=1\linewidth]{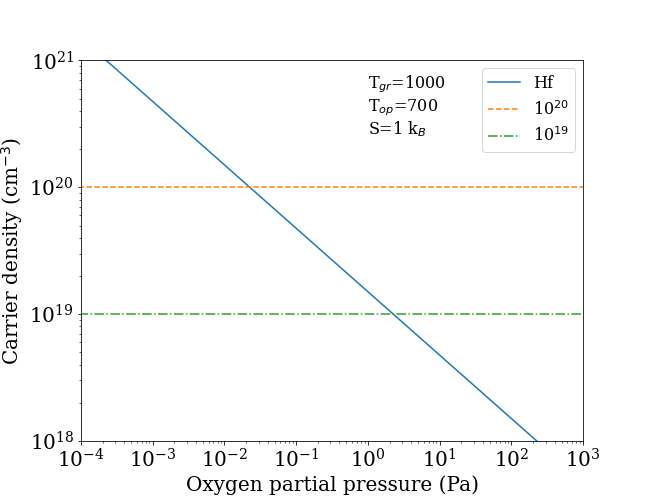}}
\caption{\label{figdop} Carrier density vs. oxygen partial pressure for Hf doping (see text for details).}
\end{figure}

To present  synthetically the results  as a function of an operational quantity, we report in Fig.~\ref{figdop} the density of carriers vs. the oxygen partial pressure at a typical  device operating temperature 
$T_{\rm op}$=700 K, and with Hf incorporated in LaSO in thermal equilibrium during growth at  temperature $T_{\rm gr}$. The rationale for this is  that, as we now discuss in more detail, the partial pressure is related to the formation energy of the donor via the oxygen chemical potential, hence determines the doping level.

\noindent
More specifically, the carrier density $N_c$ at $T_{\rm op}$ is computed as
\begin{equation} N_c=N_d  \exp{(-{\eta}/k_B T_{\rm op})}\label{Nc},\end{equation}
originating from a dopant thermal level at energy $\eta$  below to the conduction edge, with a dopant density 
\begin{equation} N_d = N_{s}\exp{(-E_{\rm f}/k_B T_{\rm gr} + S)}
\end{equation} 
embedded in thermodynamical equilibrium at the growth temperature $T_{\rm gr}$ (we assume $T_{\rm gr}$=1000 K). The   vibrational formation entropy  is arbitrarily but conservatively set to $S$=1 $k_B$, and there are $N_s$=1.697$\times$10$^{22}$ cm$^{-3}$ available sites.  For a normal dopant having a level below the band edge, $\eta$=$-\varepsilon$ (Table \ref{tab}), but since  the Hf  level is  above the conduction edge, i.e. $\varepsilon$ is positive, the ionization Arrhenius factor in Eq.~\ref{Nc} is simply set to 1.  

The formation energy is related to the oxygen chemical potential, hence to the oxygen partial pressure.   The dominant solubility limit for Zr and Hf is the formation of dioxides. Since La forms a sesquioxide, the Hf substitution causes an excess of oxygen. Therefore the formation energy $E_f$  increases with the oxygen chemical potential, and it is at its largest in oxygen-rich conditions, namely 
\begin{equation} 
E_{f} =  C + \frac{(\mu_{\rm O} - \mu_{\rm O_2})}{2},
\end{equation} 
with $\mu_{\rm O}$$\leq$$\mu_{\rm O_2}$ and 
$$C= E_{\rm def}  - E_{\rm bulk} +  \mu_{\rm La_{bulk}} - \mu_{\rm Hf_{bulk}}
 -\Delta H_{\rm HfO_2} + \Delta H_{\rm La_2O_3},$$
where $E_{\rm def}$  and $E_{\rm bulk}$ are the defected and pristine supercell energies, the $\mu$'s the chemical potentials of the bulk elements, $\mu_{\rm O_2}$ the chemical potential of O  in a O$_2$ molecule, and $\Delta H$ the oxide formation enthalpies.  {$E_{\rm def}$ and  $E_{\rm bulk}$ are calculated directly, the} bulk chemical potentials are taken from experiment, and formation enthalpies are from the Materials Project \cite{mp}. Finally, the O partial pressure   at growth  temperature $T_{\rm gr}$ is related to the  oxygen chemical potential $\mu_{\rm O}$  by the standard relation $p$=$p_0\,\exp{((\mu_{\rm O}-\mu_{\rm O_2})/k_BT_{gr})}$ and $p_0$=19 kPa the normal-conditions oxygen partial pressure. 

The dopant density  increases as the chemical potential goes more and more negative, i.e.  the partial pressure is reduced, and this offers some leeway to increase the density by adopting O-{lean} growth conditions. 
As Fig. \ref{figdop} shows, Hf can indeed produce useful carrier densities in  the   10$^{20}$ cm$^{-3}$ range for low but not unreasonable O partial pressures.  {Since the O vacancy is most likely a deep donor as it is in most oxides, oxygen deficiency should not cause notable counterdoping}. 
Zr, in turn, is unfortunately ruled out by its comparatively larger formation energy.

\section{Summary}
We  predicted a giant thermoelectric figure of merit in high-valley-multiplicity lanthanum oxysulphate LaSO. The GGA ab initio band structure, interpolated over a fine grid,  is fed into Bloch-Boltzmann theory, accounting for an energy- and temperature-dependent relaxation time   \cite{bt2} (code  available on line  \cite{thermocasu}). The lattice thermal conductivity is obtained from a model using the ab-initio phonon dispersion and Gr\"uneisen parameters, and the Debye temperature. For the perfect crystal, $ZT$ is practically linear in $T$ and at 1100 K  reaches a value between 3.5 and 6.5 depending on the lattice thermal conductivity. The optimal doping is weakly temperature dependent and in the low to mid 10$^{20}$ cm$^{-3}$ range. Our results for the power factor confirm earlier  suggestions  \cite{gibbs} that high valley multiplicity leads to large power factors and therefore large $ZT$. The $n$-type dopability of LaSO was also analyzed, suggesting Hf as a potential dopant. 

\section*{Acknowledgments}
 F.R. and G.-M.R. acknowledge support from the "Low Cost ThermoElectric Devices" (LOCOTED)
project funded by the R\'egion Wallonne (Programmes FEDER) and from CISM and CECI for computational support. 
VF, RF, and GC thank CINECA for ISCRA supercomputing grants. RF thanks ICMN-UCL for hospitality. VF is 
 on secondment leave at the Italian Embassy to Germany; his views as expressed herein are not 
necessarily shared by the Italian Ministry of Foreign Affairs.


\begin{thebibliography}{99}
\bibitem{genref}
 T. M. Tritt and M. A. Subramanian, MRS Bull. {\bf 31}, 188 (2006).

\bibitem{gibbs}
Z. M. Gibbs, F. Ricci, G. Li, H. Zhu, K. Persson, G. Ceder, G. Hautier, A. Jain, and G. J. Snyder, 
npj Comput. Mater. {\bf 3}, 8 (2017).

\bibitem{allen}
P. B. Allen, Phys. Rev. B {\bf  17}, 3725 (1978);
P. B. Allen, W. Pickett, and H. Krakauer, {\it ibid.} {\bf 37}, 7482 (1988).

\bibitem{bt2}
G. K. H. Madsen, J. Carrete, and M. J. Verstraete, Comp. Phys. Commun. {\bf 231}, 140 (2018).

\bibitem{sno}
G. Casu, A. Bosin, and V. Fiorentini, 
Phys. Rev. Materials, in print (2020)

\bibitem{noi}
R. Farris, M. B. Maccioni, A. Filippetti, and V. Fiorentini, J. Phys.: Condens. Matter {\bf 31}, 065702 (2019); M. B. Maccioni, R. Farris, and V. Fiorentini, Phys. Rev. B {\bf 98}, 220301(R) (2018).


\bibitem{thermocasu}
Available on Gitlab at {\tt http://tiny.cc/houqkz} 


\bibitem{pbe}
J. P. Perdew, K. Burke, and M. Ernzerhof,
Phys. Rev. Lett. {\bf 77}, 3865 (1996).


 \bibitem{paw}
 P. E. Bl\"ochl, Phys. Rev. B {\bf {50}}, 17953 (1994); G. Kresse and D. Joubert, 
 %%Phys. Rev. B
 {\it ibid.}, {\bf {59}}, 1758 (1999).

\bibitem{vasp}
G. Kresse and J. Furthm\"uller, Phys. Rev. B {\bf 54}, 11169 (1996);
G. Kresse and D. Joubert, {\it ibid}. {\bf 59}, 1758 (1999).



\bibitem{antimonides}
L. Bjerg,  B. B. Iversen, and G. K. H. Madsen, Phys. Rev. B {\bf 89}, 024304 (2014).


\bibitem{QE}
P. Giannozzi {\it et al.}, J. Phys.:Condens. Matter {\bf 21}, 395502 (2009); J. Phys.:Condens. Matter {\bf 29}, 465901 (2017).



\bibitem{passler}
R.  P\"assler, J. Appl. Phys. {\bf 101}, 093513 (2007).

\bibitem{ridley2}
B. K. Ridley, J. Phys.: Condens. Matter {\bf 10}, 6717 (1998).



 
\bibitem{ridley1}
B. K. Ridley, {\it Quantum Processes in Semiconductors} (Clarendon Press, Oxford, 1988).


\bibitem{bilbao}
M. I. Aroyo, J. M. Perez-Mato, D. Orobengoa, E. Tasci, G. de la Flor, and A. Kirov,
Bulg. Chem. Commun. {\bf 43}, 183 (2011); J. M. Perez-Mato, S. V. Gallego, E. S. Tasci, L. Elcoro, G. de la Flor, and M. I. Aroyo,
Annu. Rev. Mater. Res. {\bf 45}, 13.1 (2015);  Bilbao crystallographic server: {\tt http://www.cryst.ehu.es}.



\bibitem{LG}
M. Leslie and M. J. Gillan, J. Phys. C: Solid State Phys. {\bf 18}, 973 (1985).


\bibitem{hse}
 J. Heyd, G. E. Scuseria, and M. Ernzerhof, J. Chem. Phys. {\bf 118}, 8207 (2003).
%\bibitem{nye}
%J. F. Nye, {\it Physical properties of crystals} (Clarendon Press, Oxford 1985).

\bibitem{FB}
V. Fiorentini and A. Baldereschi, Phys. Rev. B {\bf 51}, 17196 (1995).
\bibitem{mp}
A. Jain, S.P. Ong, G. Hautier, W. Chen, W.D. Richards, S. Dacek, S. Cholia, D. Gunter, D. Skinner, G. Ceder, and K.A. Persson,
APL Materials {\bf 1}, 011002 (2013); {\tt https://materialsproject.org}

\end{thebibliography}
\end{document}